\documentclass[english]{article}
\usepackage{ae,aecompl}
\usepackage[T1]{fontenc}
\usepackage[latin9]{inputenc}
\usepackage{geometry}
\usepackage{babel}
\usepackage{float}
\usepackage{mathtools}
\usepackage{amsmath}
\usepackage{amssymb}
\usepackage{graphicx}
\usepackage{microtype}
\usepackage[unicode=true,
 bookmarks=true,bookmarksnumbered=false,bookmarksopen=false,
 breaklinks=false,pdfborder={0 0 0},pdfborderstyle={},backref=false,colorlinks=false]
 {hyperref}
\hypersetup{pdftitle={Optimization based evaluation of grating interferometric phase stepping series and analysis of mechanical setup instabilities},
 pdfauthor={Jonas Dittmann, Andreas Balles, Simon Zabler}}

\makeatletter

\providecommand{\tabularnewline}{\\}
\floatstyle{ruled}
\newfloat{algorithm}{tbp}{loa}
\providecommand{\algorithmname}{Algorithm}
\floatname{algorithm}{\protect\algorithmname}

\usepackage{algpseudocode}

\usepackage{fancyhdr}
\usepackage{authblk}

\makeatother

\begin{document}

\title{Optimization based evaluation of grating interferometric phase stepping
series and analysis of mechanical setup instabilities}

\author{Jonas Dittmann$^{1}$, Andreas Balles$^{1}$ and Simon Zabler$^{1,2}$\\
{\small{}$^{1}$Lehrstuhl für Röntgenmikroskopie, University of Würzburg,
Germany}\\
{\small{}$^{2}$Fraunhofer EZRT, Würzburg, Germany}}

\date{April 26, 2018}
\maketitle
\begin{abstract}
 The diffraction contrast modalities accessible by X-ray grating
interferometers are not imaged directly but have to be inferred from
sine like signal variations occurring in a series of images acquired
at varying relative positions of the interferometer's gratings. The
absolute spatial translations involved in the acquisition of these
phase stepping series usually lie in the range of only a few hundred
nanometers, wherefore positioning errors as small as 10nm will already
translate into signal uncertainties of one to ten percent in the final
images if not accounted for.

Classically, the relative grating positions in the phase stepping
series are considered input parameters to the analysis and are, for
the Fast Fourier Transform that is typically employed, required to
be equidistantly distributed over multiples of the gratings' period.

In the following, a fast converging optimization scheme is presented
simultaneously determining the phase stepping curves' parameters as
well as the actually performed motions of the stepped grating, including
also erroneous rotational motions which are commonly neglected. While
the correction of solely the translational errors along the stepping
direction is found to be sufficient with regard to the reduction of
image artifacts, the possibility to also detect minute rotations about
all axes proves to be a valuable tool for system calibration and monitoring.
The simplicity of the provided algorithm, in particular when only
considering translational errors, makes it well suitable as a standard
evaluation procedure also for large image series.

\end{abstract}
\thispagestyle{fancy}

\section{Introduction}

X-ray grating interferometry \cite{David2002,Momose2003} facilitates
access to new contrast modalities in laboratory X-ray imaging setups
and has by now been implemented by many research groups after the
seminal publication by Pfeiffer et al.\ in 2006 \cite{Pfeiffer2006}.
The additional information on X-ray refraction (``differential phase
contrast'') and ultra small angle scattering (``darkfield contrast'')
properties of a sample that can be obtained promises both increased
sensitivity to subtle material variations as well as insights into
the samples' substructure below the spatial resolution of the acquired
images.

\begin{figure}
\begin{centering}
\includegraphics[width=1\textwidth]{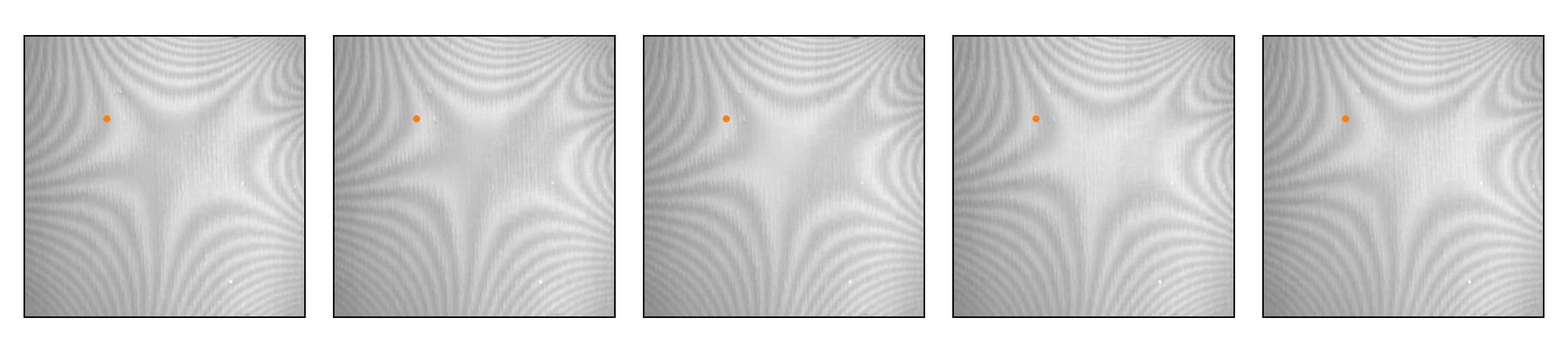}
\par\end{centering}
\begin{centering}
\includegraphics[width=0.9\textwidth]{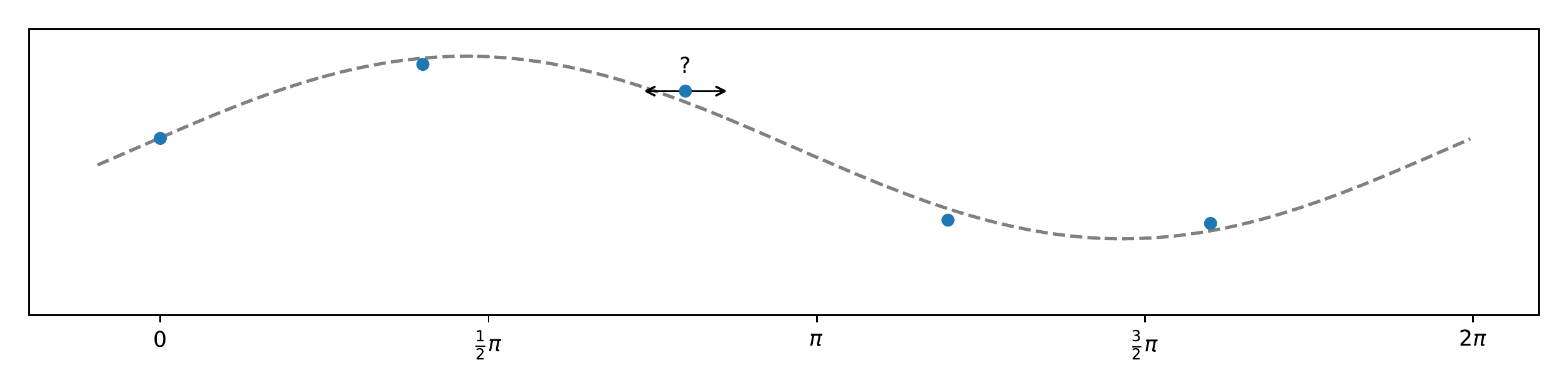}
\par\end{centering}
\centering{}\caption{\label{fig:phase-stepping-series}Example for a grating interferometric
phase stepping series. The intensity variation throughout the series
is visually most perceivable in the center. The spatial Moiré fringes
are due to imperfectly matched gratings and will translate to reference
offset phases of the sinusoid curves found at each detector pixel
(cf.\ Fig.~\ref{fig:trans-vis-ph-ref}). The bottom row shows a
corresponding phase stepping curve for the pixel marked orange in
the above image series. The sampling positions are subject to an unknown
error.}
\end{figure}

\begin{figure}
\begin{centering}
\par\end{centering}
\centering{}\includegraphics[width=0.9\textwidth]{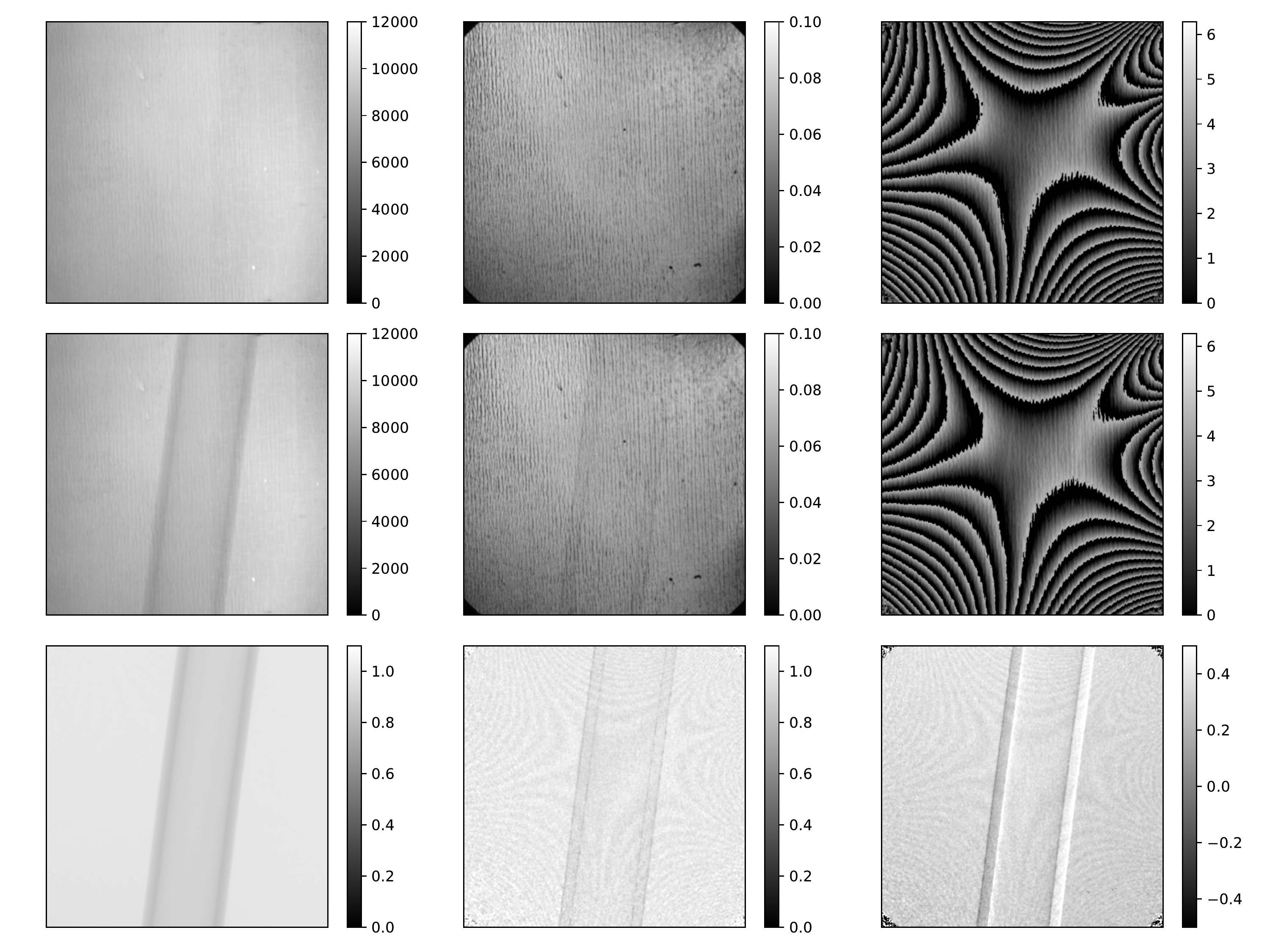}\caption{\label{fig:trans-vis-ph-ref}From left to right: transmission (sinusoid
mean), visibility (ratio of sinusoid amplitude and mean) and phase
images derived from phase stepping series as shown in Fig.~\ref{fig:phase-stepping-series}.
The first rows show acquisitions with and without sample, while the
last row shows the sample images (center row) normalized with respect
to the empty beam images (first row). Positioning errors in the phase
stepping procedure cause the Moiré pattern of the reference phase
image to translate into the final results.}
\end{figure}

In contrast to classic X-ray imaging, the absorption, differential
phase and darkfield contrasts are not imaged directly but are encoded
in sinusoidal intensity variations arising at each detector pixel
when shifting the interferometer's gratings relative to each other
perpendicular to the beam path and the grating bars. A crucial step
in the generation of respective absorption, phase and darkfield images
therefore is the analysis of the commonly acquired phase stepping
series, which shall be the subject of the present article. Respective
examples are shown in Figures \ref{fig:phase-stepping-series} and
\ref{fig:trans-vis-ph-ref}.

In principle, the images within such phase stepping series are sampled
at about five to ten different relative grating positions equidistantly
distributed over multiples of the gratings' period such that the expected
sinusoids for each detector pixel may be characterized by standard
Fourier decomposition. The zeroth order term represents the mean transmitted
intensity (as in classic X-ray imaging), while the first order terms
encode phase shift and amplitude of the sinusoid. The ratio of amplitude
and mean (generally referred to as ``visibility'') is here related
to scattering and provides the darkfield contrast. Higher order terms
correspond to deviations from the sinusoid model mainly due to the
actual grating profiles and are usually not considered.

Given typical grating periods in the range of two to ten micrometers,
the actually performed spatial translations lie in the range of 200
to 2000 nanometers. Particular for the smaller gratings, positioning
errors as small as 10 to 20 nm imply relative phase errors in the
range of five to ten percent, causing uncertainties in the derived
quantities in the same order of magnitude. The propagation of noise
within the sampling positions onto the extracted signals has e.g.\
been studied by Revol et al.\ \cite{Revol2010} and first results
for the determination of the actual sampling positions from the available
image series were shown by Seifert et al.\ \cite{Seifert2016} using
methods by Vargas et al.\ \cite{Vargas2013} from the context of
visible light interferometry. Otherwise the problem seems to have
not been given much consideration so far.

The present article proposes a simple iterative optimization algorithm
both for the fitting of irregularly sampled sinusoids and in particular
also for the determination of the actual sampling positions. The use
of only basic mathematical operations eases straightforward implementations
on arbitrary platforms. Besides uncertainties in the lateral stepping
motion, the remaining mechanical degrees of freedom (magnification/expansion
and rotations) are also considered. The proposed techniques will be
demonstrated on an exemplary data set. 

\section{Methods}

The task of sinusoid fitting with imprecisely known sampling locations
will be partitioned into two separate optimization problems considering
either only the sinusoid parameters or only the sampling positions
while temporarily fixing the respective other set of parameters. Alternating
both optimization tasks will minimize the joint objective function
in few iterations.

\subsection{Sinusoid Fitting}

A simple iterative forward-backprojection algorithm converging to
the least squares solution can be derived when representing the sinusoid
model as a linear combination of basis functions:
\begin{equation}
\begin{aligned}o+a\sin(\phi-\phi_{0}) & =o+a\left(\cos\phi_{0}\sin\text{\ensuremath{\phi}}-\sin\phi_{0}\cos\phi\right)\\
 & =o+a_{s}\sin\phi+a_{c}\cos\phi\\
 & =(o,a_{s},a_{c})\cdot(1,\sin\phi,\cos\phi)^{T}
\end{aligned}
\label{eq:sinusoid-model}
\end{equation}
with the following identities:
\begin{equation}
\begin{aligned}a & =\sqrt{a_{s}^{2}+a_{c}^{2}}\\
\sin\phi_{0} & =-a_{c}/a\\
\cos\phi_{0} & =a_{s}/a\\
\phi_{0} & =\arctan\!2(-a_{c},a_{s})\,.
\end{aligned}
\label{eq:sinusoid-param-conversions}
\end{equation}
The components $o$, $a_{c}$, $a_{s}$ may be determined by means
of the following iterative scheme (introducing the sample index $i$
and the superscript iteration index $k$):
\begin{equation}
\begin{aligned}o^{(0)},a_{c}^{(0)},a_{s}^{(0)} & =0,0,0\\
\tilde{y}_{i}^{(k)} & =o^{(k)}+a_{c}^{(k)}\cos\phi_{i}+a_{s}^{(k)}\sin\phi_{i}\\
o^{(k+1)} & =o^{(k)}+\frac{1}{N}\sum_{i}(y_{i}-\tilde{y}_{i}^{(k)})\\
a_{s}^{(k+1)} & =a_{s}^{(k)}+\frac{2}{N}\sum_{i}(y_{i}-\tilde{y}_{i}^{(k)})\sin\phi_{i}\\
a_{c}^{(k+1)} & =a_{c}^{(k)}+\frac{2}{N}\sum_{i}(y_{i}-\tilde{y}_{i}^{(k)})\cos\phi_{i}
\end{aligned}
\label{eq:iterative-sine-fitting}
\end{equation}
where the factors of $1/N$ and $2/N$ account for the normalization
of the respective basis functions (with $N$ being the amount of samples
$(\phi_{i},y_{i})$ enumerated by $i$). The scheme reduces to classic
Fourier analysis for the case of the abscissas $\phi_{i}$ being equidistantly
distributed over multiples of $2\pi$ and converges within the first
iteration in that case. As the update terms to $o$, $a_{c}$ and
$a_{s}$ are proportional to the respective derivatives of the $\ell_{2}$
error $\sum_{i}(o^{(k)}+\,a_{c}^{(k)}\!\cos\phi_{i}\,+\,a_{s}^{(k)}\!\sin\phi_{i}\,-y_{i})^{2}$,
the fixpoint of the iteration will be the least squares fit also in
all other cases.

For a stopping criterion, the relative error reduction 
\begin{equation}
\Delta_{\ell_{2}}=\frac{\sqrt{\sum_{i}^{N}\left(y_{i}-\tilde{y}_{i}^{(k-1)}\right)^{2}}-\sqrt{\sum_{i}^{N}\left(y_{i}-\tilde{y}_{i}^{(k)}\right)^{2}}}{\sqrt{\sum_{i}^{N}\left(y_{i}-\tilde{y}_{i}^{(k)}\right)^{2}}}\label{eq:l2-error-reduction}
\end{equation}
may be tracked. It is typically found to fall below 0.1\% within 10
to 20 iterations given only slightly noisy data (noise sigma three
orders of magnitude smaller than sinusoid amplitude) and within less
than 10 iterations for most practical cases. For the special case
of equidistributed $\phi_{i}$ on multiples of $2\pi$, it will immediately
drop to 0 after the first iteration. In practice, a fixed amount of
iterations in the range of five to fifteen will therefore be adequate
as stopping criterion as well.

\subsection{Phase Step Optimization}

An underlying assumption of the previously described least squares
fitting procedure is the certainty of the abscissas, i.e.\ the set
of phases $\phi_{i}$ at which the ordinates $y_{i}$ have been sampled.
As the sampling positions are themselves subject to experimental uncertainties
(arising from the mechanical precision of the involved actuators),
a further optimization step will be introduced that minimizes the
least squares error of the sinusoid fit over the sampling positions
$\phi_{i}$. While this procedure obviously results in overfitting
when considering only a single phase stepping curve (PSC), it becomes
a well-defined error minimization problem when regarding large sets
of PSCs sharing the same $\phi_{i}$. In other words, an approach
to the minimization of the objective function
\begin{equation}
o_{j},a_{j},\phi_{0,j},\Delta\phi_{i}=\underset{o_{j},a_{j},\phi_{0,j},\Delta\phi_{i}}{\text{argmin}}\sum_{i,j}\left(o_{j}+a_{j}\sin(\phi_{i}+\Delta\phi_{i}-\phi_{0_{j}})-y_{ij}\right)^{2}\label{eq:global-error-min-task}
\end{equation}
shall be considered, where $j$ indexes detector pixels.

In order to derive an optimization procedure for the sampling positions,
first the fictive case of a perfectly sinusoid PSC with negligible
statistical error on the ordinate (the sampled values) shall be considered.
Ignoring for now the fact that least squares fits commonly assume
only the ordinates to be affected by noise, a least squares fit shall
be used to preliminarily determine the parameters of the sinusoid
described by the observed data. Assuming then that inconsistencies
of the observed data with the model are due to errors on the sampling
locations, deviations from their intended positions are given by the
data points' lateral distances from the sinusoid curve (cf.\ Fig.~\ref{fig:delta-phi-demo}).
Finally, the actual systematic deviations of the sampling locations
can be found by averaging over the respective results for a large
set of PSCs sampled simultaneously. This information can be fed back
into the original sinusoid fit, which then again allows the refinement
of the current estimate of the true sampling positions, finally resulting
in an iterative procedure alternatingly optimizing the sinusoid parameters
and the actual sampling locations (cf.\ Algorithm~\ref{alg:sinusoid-abscissa-optimisation}). 

\subsubsection{Determination of individual phase deviations}

\begin{figure}
\centering{}\includegraphics[width=0.45\textwidth]{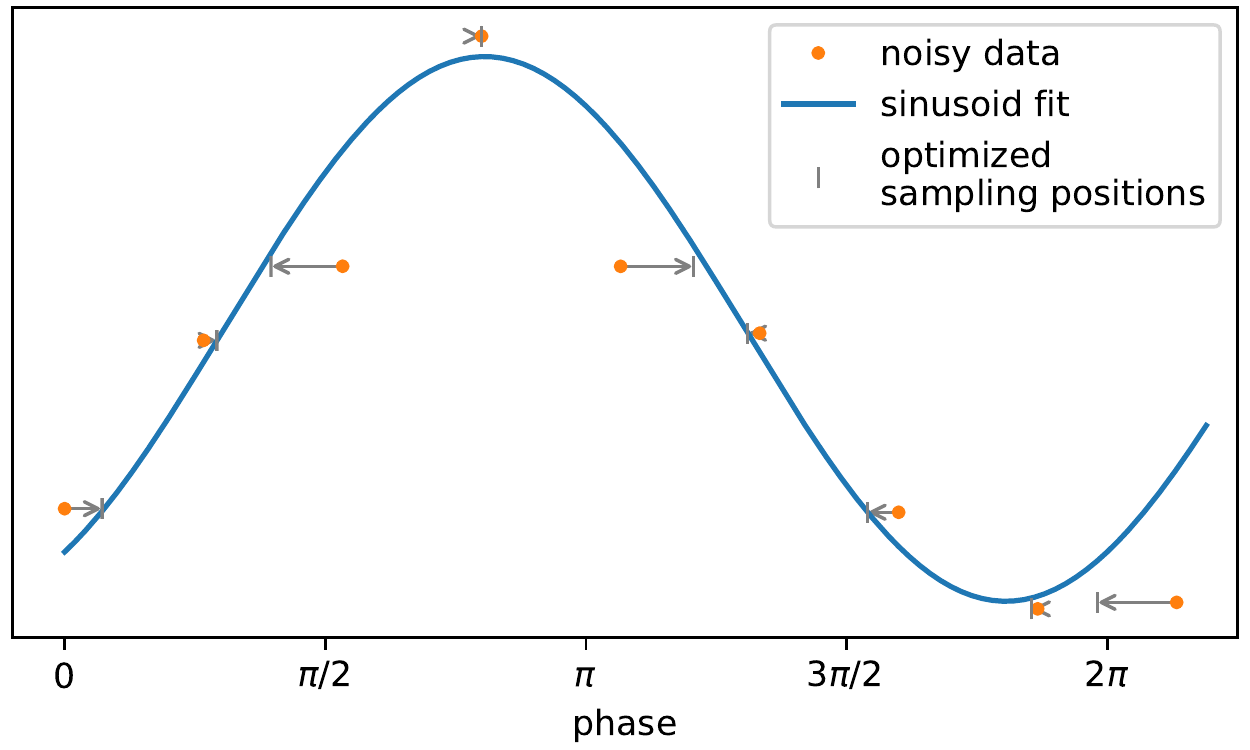}\caption{\label{fig:delta-phi-demo}Optimization of the actual (in contrast
to the intended) sampling positions by Eq.~\ref{eq:delta-phi} with
respect to a previous sinusoid fit based on the temporary assumption
that deviations from the sinusoid model are mainly due to errors on
the sampling positions rather than the sampled ordinate values. Averaging
of respective phase deviations found for large sets of measurements
will finally allow the differentiation of systematic deviations from
statistical noise (see also Fig.~\ref{fig:dphi-and-w-maps}).}
\end{figure}
Starting with an initial sinusoid fit 
\begin{equation}
o,a,\phi_{0}=\underset{o,a,\phi_{0}}{\text{argmin}}\sum_{i}\left(o+a\sin(\phi_{i}-\phi_{0})-y_{i}\right)^{2}\,,
\end{equation}
the error shall be minimized over deviations $\Delta\phi_{i}$ to
$\phi_{i}$ while keeping the sinusoid model parameters fixed:

\begin{equation}
\Delta\phi_{i}=\underset{\Delta\phi_{i}}{\text{argmin}}\sum_{i}\left(o+a\sin(\phi_{i}+\Delta\phi_{i}-\phi_{0})-y_{i}\right)^{2}\,.\label{eq:phi-optimization-task}
\end{equation}

\begin{figure}
\centering{}\includegraphics[width=0.45\textwidth]{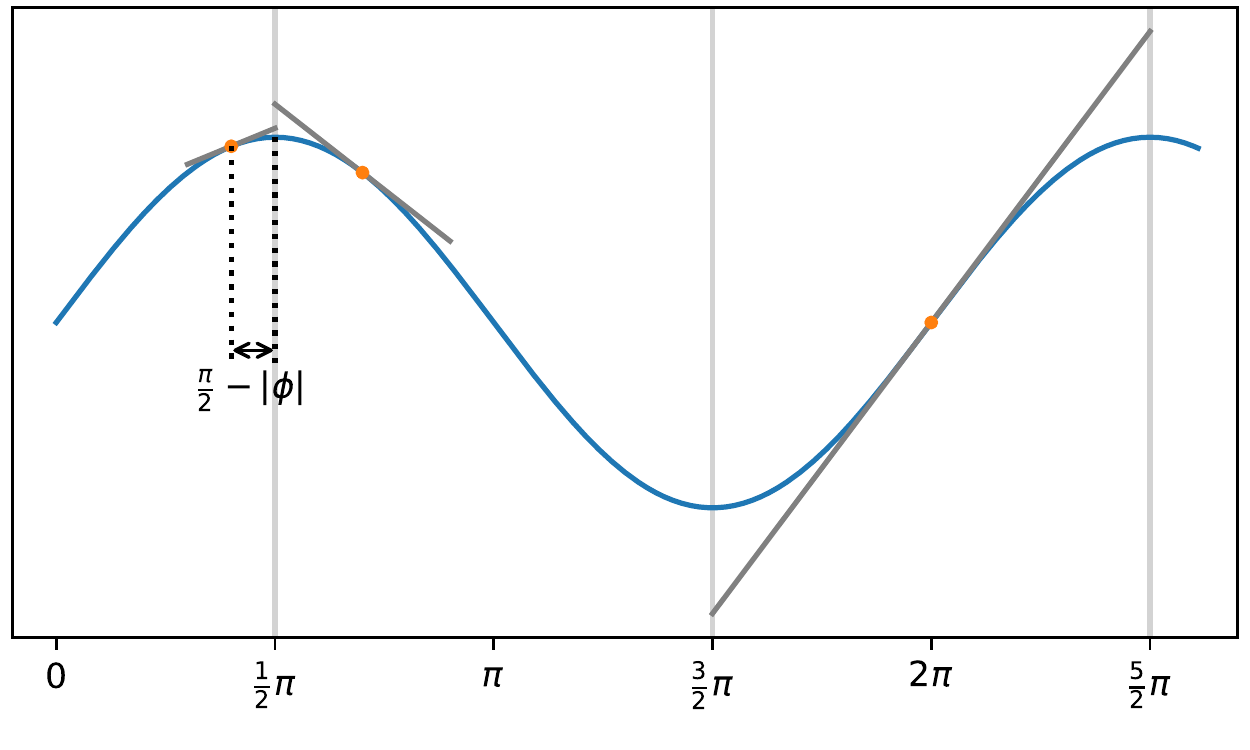}\caption{\label{fig:taylor-limits}Illustration of the maximal sensible ranges
for linear approximations of a sinusoid. Vertical lines at the turning
points indicate the boundaries of monotone sections that should never
be crossed by linear approximations of the curve. The maximum meaningful
range is thus largest for points furthest from these boundaries and
reduces to zero exactly at the turning points. Equation~\ref{eq:taylor-validity-range}
approximates this phase dependence of the validity range with a $\cos^{2}$
function, and Eq.~\ref{eq:m0-max} defines the maximum amplitude
admissible in order to indeed never exceed turning points.}
\end{figure}

Equation~\ref{eq:phi-optimization-task} is solved when the derivative
of the objective function with respect to $\Delta\phi_{i}$ vanishes:
\begin{equation}
\begin{aligned}0 & =\frac{\text{d}}{\text{d}\Delta\phi_{i}}\sum_{i}\left(o+a\sin(\phi_{i}+\Delta\phi_{i}-\phi_{0})-y_{i}\right)^{2}\\
0 & =2\left(o-y_{i}+a\sin(\phi_{i}+\Delta\phi_{i}-\phi_{0})\right)a\cos(\phi_{i}+\Delta\phi_{i}-\phi_{0})\\
 & \text{for }a\cos(\phi_{i}+\Delta\phi_{i}-\phi_{0})\neq0:\\
0 & =\left(o-y_{i}+a\sin(\phi_{i}+\Delta\phi_{i}-\phi_{0})\right)\\
 & \text{for }\Delta\phi_{i}\ll\pi:\\
0 & \approx\left(o-y_{i}+a\sin(\phi_{i}-\phi_{0})+\Delta\phi_{i}a\cos(\phi_{i}-\phi_{0})\right)
\end{aligned}
\label{eq:dphi-approximate-l2-min}
\end{equation}
where the last step is a first order Taylor expansion with respect
to $\Delta\phi_{i}$. This directly leads to the following expression
for $\Delta\phi_{i}$:
\begin{equation}
\Delta\phi_{i}\approx\frac{1}{\cos(\phi_{i}-\phi_{0})}\left(\frac{y_{i}-o}{a}-\sin(\phi_{i}-\phi_{0})\right)\quad\quad\text{for }\Delta\phi_{i}\ll\pi\text{ and }a\cos(\phi_{i}-\phi_{0})\neq0\,\,,
\end{equation}
where the earlier condition $a\cos(\phi_{i}+\Delta\phi_{i}-\phi_{0})\neq0$
is approximated to be satisfied when $a\cos(\phi_{i}-\phi_{0})\neq0$.
The restriction to cases with $a\cos(\phi_{i}-\phi_{0})\neq0$ can
be intuitively understood when recalling that $\cos(\phi_{i}-\phi_{0})=0$
implies a maximum or minimum of the sinusoid and $a=0$ means that
it is constant ($\phi_{i}$ independent), in both of which cases there
is no sensible choice for $\Delta\phi_{i}\neq0$. The constraint on
the result, $\Delta\phi_{i}\ll\pi$, can simply be taken into account
by means of a limiting function such as
\begin{align}
\text{softlimit}(\Delta\phi_{i},m\geq0) & =\begin{cases}
0 & m=0\\
m\tanh\left(\frac{\Delta\phi_{i}}{m}\right) & \text{else}
\end{cases}\,\,,\label{eq:softlimit}
\end{align}
which provides a linear mapping with slope $1$ for $\Delta\phi_{i}\ll m$
and is bounded at $\pm m$. The choice of $m$ in this case depends
on the validity range of the linear approximation of $\sin(\phi_{i}+\Delta\phi_{i}-\phi_{0})$
with respect to $\Delta\phi_{i}$ about $\phi_{i}-\phi_{0}$, which
obviously depends on the magnitude of the curvature of the sinusoid
at this point as illustrated in Figure~\ref{fig:taylor-limits}.
The latter may be accounted for by 
\begin{equation}
m=m_{\phi_{i}-\phi_{0}}=m_{0}\cos^{2}(\phi_{i}-\phi_{0})\label{eq:taylor-validity-range}
\end{equation}
which reaches its maximum $m_{\phi_{i}-\phi_{0}}=m_{0}$ at $\phi_{i}-\phi_{0}=0$
(where $\sin(\phi_{i}-\phi_{0})$ is actually linear) and smoothly
drops to $0$ for $\cos(\phi_{i}-\phi_{0})=0$, in which case both
the sine and its curvature are maximal and $\Delta\phi_{i}$ shall
remain $0$. The actual choice of $m_{0}$ can now be based on the
validity range of the linear approximation about $\phi_{i}-\phi_{0}=0$.
The upper bound to $m_{\phi_{i}-\phi_{0}}$ and thus to $m_{0}$ is
defined by the range over which $\sin(x-x_{0}),\,x\in[-m_{0}\cos^{2}(x_{0}),+m_{0}\cos^{2}(x_{0})]$
is actually invertible (cf.\ Fig.~\ref{fig:taylor-limits}), i.e.:
\begin{equation}
\begin{aligned}m_{0}\cos^{2}(\phi) & \leq\frac{\pi}{2}-|\phi|\\
m_{0} & \lesssim1.38\,.
\end{aligned}
\label{eq:m0-max}
\end{equation}
For $m_{0}=1.38$, the linear approximation used in Eq.~\ref{eq:dphi-approximate-l2-min}
deviates by up to $40\%$. The deviation is reduced to $20\%$ or
$5\%$ for $m_{0}=1$ and $m_{0}=\frac{1}{2}$ respectively. 

Combining the above results and introducing for completeness also
the detector pixel index $j$, the following expression for $\Delta\phi_{ji}$
reducing (and, for sufficiently small $\Delta\phi_{ji}$, minimizing)
the $\ell_{2}$ error $\left(o_{j}+a_{j}\sin(\phi_{i}+\Delta\phi_{ji}-\phi_{0,j})-y_{ji}\right)^{2}$
can be given, choosing $m_{0}=\frac{1}{2}$:
\begin{equation}
\Delta\phi_{ji}\approx\begin{cases}
0 & a\cos(\phi_{i}-\phi_{0,j})=0\\
\text{softlimit}\left(\frac{\left(\frac{y_{ji}-o_{j}}{a_{j}}-\sin(\phi_{i}-\phi_{0,j})\right)}{\cos(\phi_{i}-\phi_{0,j})},\frac{1}{2}\cos^{2}(\phi_{i}-\phi_{0,j})\right) & \text{else}
\end{cases}\,\,,\label{eq:delta-phi}
\end{equation}
Figure~\ref{fig:delta-phi-demo} shows an example of this approximate
least squares solution to $\Delta\phi_{ji}$.

\subsubsection{Noise weighted average of phase deviations\label{subsec:mean-phase-deviation}}

Now that an expression has been derived for the deviations $\Delta\phi_{ji}$
optimizing the abscissa values for a single PSC given a previous sinusoid
fit, the respective results for many PSCs (indexed by $j$) sharing
the same sampling locations may be averaged:
\begin{equation}
\Delta\phi_{i}=\frac{\sum_{j}w_{ji}\Delta\phi_{ji}}{\sum_{j}w_{ji}}\,,
\end{equation}
using weights $w_{ji}$ factoring in the relative certainty and relevance
of the different $\Delta\phi_{ji}$. An appropriate choice is
\begin{equation}
w_{ji}=a_{j}^{2}\cos^{2}(\phi_{i}-\phi_{0,j})\,,\label{eq:delta-phi-importance-w}
\end{equation}
where $\cos^{2}(\phi_{i}-\phi_{0,j})$ weights the slope dependent
error propagation from noisy measurements $y_{ji}$ to $\Delta\phi_{ji}$
based on the derivative of the sinusoid model at $\phi_{i}$ and $a_{j}^{2}$
weights the contribution of a particular PSC $j$ to the accumulated
$\ell_{2}$ error. These considerations lead to 
\begin{equation}
\begin{aligned}\Delta\phi_{i} & =\frac{\sum_{j}a_{j}^{2}\cos^{2}(\phi_{i}-\phi_{0,j})\Delta\phi_{ji}}{\sum_{j}a_{j}^{2}\cos^{2}(\phi_{i}-\phi_{0,j})}\\
 & =\frac{\sum_{j}a_{j}^{2}\cos^{2}(\phi_{i}-\phi_{0,j})\text{softlimit}\left(\frac{\left(\frac{y_{ji}-o_{j}}{a_{j}}-\sin(\phi_{i}-\phi_{0,j})\right)}{\cos(\phi_{i}-\phi_{0,j})},\frac{1}{2}\cos^{2}(\phi_{i}-\phi_{0,j})\right)}{\sum_{j}a_{j}^{2}\cos^{2}(\phi_{i}-\phi_{0,j})}\\
 & =\frac{\sum_{j}\text{softlimit}\left(\cos(\phi_{i}-\phi_{0,j})\left(a_{j}(y_{ji}-o_{j})-a_{j}^{2}\sin(\phi_{i}-\phi_{0,j})\right),\frac{1}{2}a_{j}^{2}\cos^{4}(\phi_{i}-\phi_{0,j})\right)}{\sum_{j}a_{j}^{2}\cos^{2}(\phi_{i}-\phi_{0,j})}\,\,,
\end{aligned}
\label{eq:delta-phi-average}
\end{equation}
where the last step uses the relation $\alpha\,\text{softlimit}(x,m)=\text{softlimit}(\alpha x,\alpha m)$
for $\alpha\geq0$.

Finally, the above derivations can be combined to an iterative optimization
algorithm reducing the accumulated least square error of multiple
sinusoid fits (indexed by $j$) to data points $y_{ji}$ over shared
abscissa values $\phi_{i}$ as defined by Equation~\ref{eq:global-error-min-task}.
A pseudo code representation is given in Alg.~\ref{alg:sinusoid-abscissa-optimisation},
further introducing the relaxation parameter $\text{\ensuremath{\lambda}}_{k}\in(0;1]$
that may be chosen $<1$ in order to damp the adaptions to $\phi_{i}^{(k)}$
if desired. The intermediate sinusoid fits may be accomplished using
the iterative algorithm described in the previous section.

\begin{algorithm}
\begin{algorithmic}[1]
\State ${\phi_{i}}, {y_{ji}}$: input data 
\State $m_{0} \gets \frac{1}{2}$ \Comment upper limit to $\Delta\phi_{ji}, m_0\in(0;1.38]$
\State $\phi_{i}^{(0)} \gets \phi_{i}$
\State $o_{j}^{(0)},a_{j}^{(0)},\phi_{0,j}^{(0)} \gets \underset{o,a,\phi_{0}}{\text{argmin}}\sum_{i}\left(o+a\sin(\phi_{i}^{(0)}-\phi_{0})-y_{ji}\right)^{2}$ \Comment initalization
\For{$k=0\,..\,k_{\max}$}

  \State $\Delta\phi_{i}^{(k)} \gets \frac{\sum_{j}\text{softlimit}\left(\cos(\phi_{i}^{(k)}-\phi_{0,j}^{(k)})\left(a_{j}^{(k)}(y_{ji}-o_{j}^{(k)})-a_{j}^{2}\sin(\phi_{i}^{(k)}-\phi_{0,j}^{(k)})\right),(a_{j}^{(k)})^{2}m_{0}\cos^{4}(\phi_{i}^{(k)}-\phi_{0,j}^{(k)})\right)}{\sum_{j}(a_{j}^{(k)})^{2}\cos^{2}(\phi_{i}^{(k)}-\phi_{0,j}^{(k)})}$

   \State $\phi_{i}^{(k+1)} \gets \phi_{i}^{(k)}+\lambda_{k}\Delta\phi_{i}^{(k)}$                   
  \State $o_{j}^{(k+1)},a_{j}^{(k+1)},\phi_{0,j}^{(k+1)} \gets \underset{o_j,a_j,\phi_{0,j}}{\text{argmin}}\sum_{i}\left(o_j+a_j\sin(\phi_{i}^{(k+1)}-\phi_{0,j})-y_{ji}\right)^{2}$

\EndFor
.
\end{algorithmic}

\caption{\label{alg:sinusoid-abscissa-optimisation}Least squares optimization
of shared abscissa values $\phi_{i}$ for simultaneous sinusoid fits
to ordinate samples $y_{ji}$ belonging to independent curves $j$
sampled at identical positions $\phi_{i}$. This represents a special
case of Algorithm~\ref{alg:sinusoid-optimization-w-gradient-model}
with spatially invariant sampling phases. The relaxation parameter
$\lambda_{k}\in(0;1]$ may be chosen $<1$ if damping of the updates
to $\phi_{i}^{(k)}$ is desired. For the intermediate argmin operations
see Equations~\ref{eq:sinusoid-model}\textendash \ref{eq:l2-error-reduction}. }
\end{algorithm}

\subsubsection{Inhomogeneous sampling phase deviations}

Up to now, it has been assumed that deviations from the intended phase
stepping positions are due to purely translational uncertainties in
the relative motion of the involved gratings, resulting in offsets
$\Delta\phi_{i}$ of the actual from the intended sampling phases
that are homogeneous throughout the whole detection area. When also
considering relative grating period changes (e.g.\ due to either
thermal expansion or motion induced changes in magnification) and
rotary motions of the interferometer's gratings relative to each other
(e.g.\ due to backlashes within the mechanical actuators), the effective
sampling phases at each phase step may exhibit gradients over the
detection area. Given the small grating periods (micrometer scale)
compared to the total extents of the gratings (centimeter scale),
both tilts in the sub-microrad range and relative period changes in
the range of $10^{-7}$ will already manifest themselves in observable
gradients.

\begin{figure}
\centering{}\includegraphics[width=0.9\textwidth]{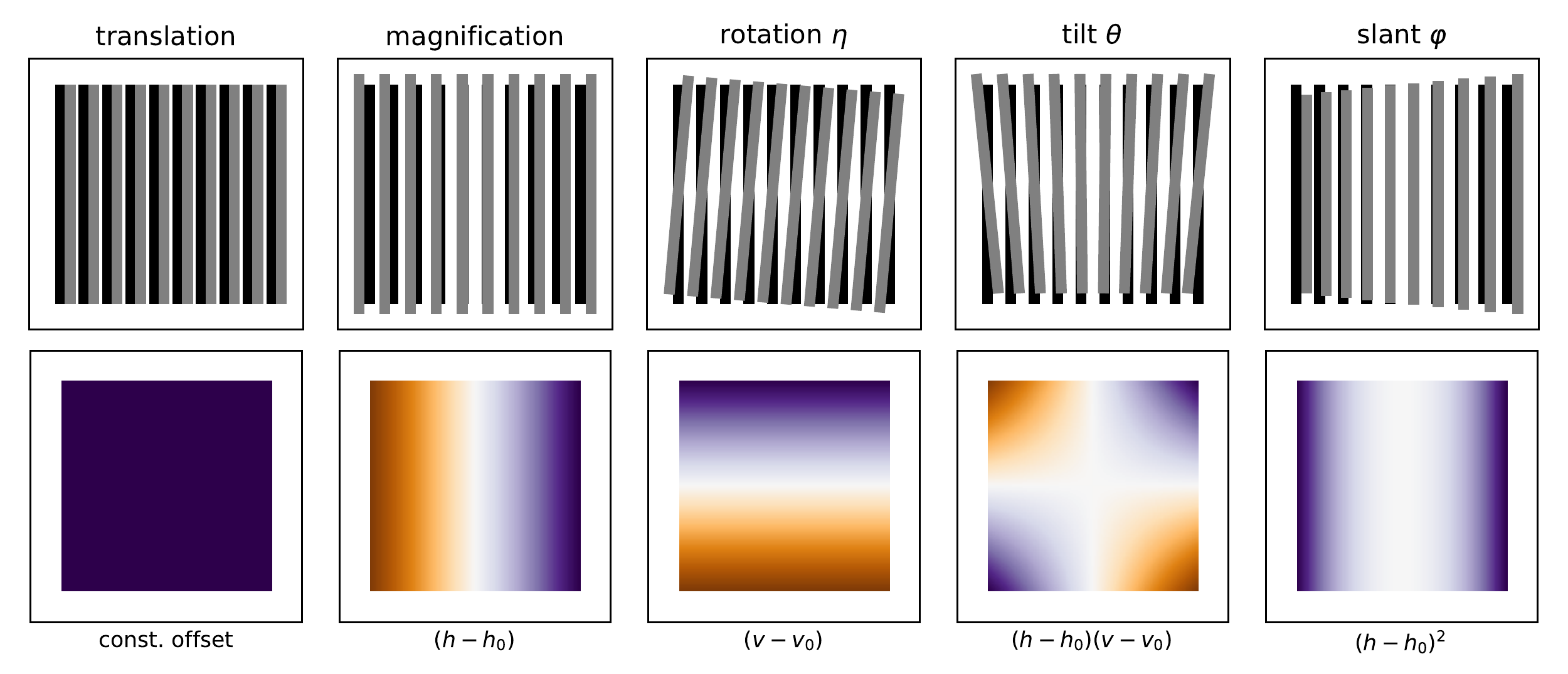}\caption{\label{fig:grating-misalignments}Grating misalignments (top row)
and corresponding spatial phase variations (bottom row). From left
to right: translation, magnification, rotation, tilt, slant. The latter
two effects (as well as translation induced magnification) are only
observable in cone beam setups. The employed colorbar ranges from
orange for negative values over white (zero) to blue for positive
values.}
\end{figure}
The corresponding optimization problem regarding also gradients is,
analog to Equation~\ref{eq:global-error-min-task}, given by:
\begin{equation}
o_{j},a_{j},\phi_{0,j},\Delta\phi_{i}(j)=\underset{o_{j},a_{j},\phi_{0,j},\Delta\phi_{i}(j)}{\text{argmin}}\sum_{i,j}\left(o_{j}+a_{j}\sin(\phi_{i}+\Delta\phi_{i}(j)-\phi_{0_{j}})-y_{ij}\right)^{2}\label{eq:global-error-min-w-gradients}
\end{equation}
where the spatial dependence of the phase deviations $\Delta\phi_{i}$
has been accounted for by a functional dependence on the detector
pixel index $j$. Given the expected gradients (as illustrated in
Figure~\ref{fig:grating-misalignments}), $\Delta\phi_{i}(j)$ has
the following form:
\begin{equation}
\Delta\phi_{i}(j)=\Delta\phi_{i}+\nabla_{\!h}\phi_{i}\,(h-h_{0})+\nabla_{\!v}\phi_{i}\,(v-v_{0})+\nabla_{\!hv}\phi_{i}(h-h_{0})(v-v_{0})+\nabla_{\!h}^{2}\phi_{i}(h-h_{0})^{2}\,,\label{eq:spatial-phasediff-model}
\end{equation}
with $\nabla_{\!h}\phi_{i}$, $\nabla_{\!v}\phi_{i}$, $\nabla_{\!hv}\phi_{i}$
and $\nabla_{\!h}^{2}\phi_{i}$ quantifying the respective gradients
in horizontal and vertical direction as well as the mixed term and
the curvature in horizontal direction, and $h$ and $v$ being spatial
detector pixel indices related to the linear pixel index $j$ through
the amount $N_{h}$ of pixels within one detector row:
\begin{equation}
\begin{aligned}j & =vN_{h}+h\\
h & =j\bmod N_{h}\\
v & =(j-h)/N_{h}\,.
\end{aligned}
\label{eq:spatial-det-indices}
\end{equation}
The constant offsets $h_{0}$ and $v_{0}$ characterize the detector
center.

The optimization of the extended objective function in Equation~\ref{eq:global-error-min-w-gradients}
can be performed analog to that of Eq.~\ref{eq:global-error-min-task}
with the only difference lying in the evaluation of the spatial phase
difference maps $\Delta\phi_{ji}$ defined by Eq.~\ref{eq:delta-phi}
(an example of which is shown in Figure~\ref{fig:dphi-and-w-maps}).
The weighted average derived in the previous section in order to determine
the homogeneous offset $\Delta\phi_{i}$ can be extended to a generalized
linear least squares fit of the model $\Delta\phi_{i}(j)=\Delta\phi_{i}(h(j),v(j))$
defined by Eq.~\ref{eq:spatial-phasediff-model} to the local estimates
$\Delta\phi_{ij}$ (Eq.~\ref{eq:delta-phi}, Fig.~\ref{fig:dphi-and-w-maps}),
also taking the weights defined by Eq.~\ref{eq:delta-phi-importance-w}
into account. Said procedure is stated more formally in Algorithm~\ref{alg:sinusoid-optimization-w-gradient-model}.

Basic geometric considerations neglecting higher order interrelations
of the considered effects (e.g.\ rotation and effective period change)
result in the following relations between the observable parameters
$\nabla_{\!h}\phi_{i}$, $\nabla_{\!v}\phi_{i}$, $\nabla_{\!hv}\phi_{i}$,
$\nabla_{\!h}^{2}\phi_{i}$ and relative translatory and rotatory
motions of the interferometer's gratings. In order to relate various
magnification changes to spatial motions based on the intercept theorem,
an assumption has to be made as to which of the gratings actually
moved. Here, the grating that is mounted on the linear phase stepping
actuator is assumed to be the cause of all relative motions of both
gratings also including tilts and rotations. The ``source\textendash grating
distance'' in the following equations will thus refer to the stepped
grating.

$\Delta\phi_{i}$ quantifies the translational error analog to Section~\ref{subsec:mean-phase-deviation}.
In contrast to the previous section, the present model distinguishes
between homogeneous phase deviations induced by translation and the
mean component induced by the $\nabla_{\!h}^{2}\phi_{i}(h-h_{0})^{2}$
term in the case of non-vanishing curvature of the spatial phase deviation\@.

The vertical gradient parameter $\nabla_{\!v}\phi_{i}$ is related
to a relative rotation $\text{\ensuremath{\eta}}$ of both gratings
about the optical axis:
\begin{align}
\tan\eta & =\frac{\nabla_{\!v}\phi_{i}}{2\pi}\frac{\text{effective grating period}}{\text{detector pixel pitch}}\,,\label{eq:grating-rotation}
\end{align}
where the ``effective grating period'' refers to the projected period
length at the location of the detector, which should be identical
for both interferometer gratings (not considering the optional additional
coherence grating close to the X-ray source).

The horizontal gradient parameter $\nabla_{\!h}\phi_{i}$ is related
to a relative mismatch in effective grating periods of the gratings
either due to relative distance changes along the optical axis or
due to actual expansions (e.g.\ thermally induced):
\begin{equation}
\text{relative period mismatch}=\frac{\text{effective period difference}}{\text{effective grating period}}=\frac{\nabla_{\!h}\phi_{i}}{2\pi}\frac{\text{effective grating period}}{\text{detector pixel pitch}}\,.
\end{equation}
When assuming relative grating period mismatches to be caused by changes
in magnification due to translations of one of the gratings along
the optical axis, the following relation applies to first order:
\begin{equation}
\begin{aligned}\text{translation distance} & =(\text{relative period mismatch})\frac{(\text{source--grating distance})^{2}}{\text{source--detector distance}}\\
 & =\frac{\nabla_{\!h}\phi_{i}}{2\pi}\frac{\text{effective grating period}}{\text{detector pixel pitch}}\frac{(\text{source--grating distance})^{2}}{\text{source--detector distance}}\,.
\end{aligned}
\end{equation}

The change $\nabla_{\!hv}\phi_{i}$ of the horizontal gradient throughout
the vertical direction corresponds to a relative change in magnification
from top to bottom, e.g.\ due to a tilt $\theta$ of one of the gratings
about the horizontal axis. Using the above relation between magnification
changes and spatial displacements, the tilt $\theta$ about the horizontal
axis is related to $\nabla_{\!hv}\phi_{i}$ approximately by
\begin{equation}
\begin{aligned}\tan\theta & =\frac{\nabla_{\!hv}\phi_{i}}{2\pi}\frac{\text{effective grating period}}{\text{detector pixel pitch}}\left((\text{detector pixel pitch})\frac{\text{source--grating distance}}{\text{source--detector distance}}\right)^{-1}\times\\
 & \quad\quad\quad\quad\quad\quad\quad\quad\quad\quad\quad\quad\quad\quad\quad\quad\quad\quad\quad\quad\quad\quad\quad\quad\quad\times\frac{(\text{source--grating distance})^{2}}{\text{source--detector distance}}\\
 & =\frac{\nabla_{\!hv}\phi_{i}}{2\pi}\frac{(\text{effective grating period})(\text{source--grating distance})}{(\text{detector pixel pitch})^{2}}\,.
\end{aligned}
\end{equation}

A non-vanishing curvature $\nabla_{\!h}^{2}\phi_{i}$ arises in case
of a rotary motion about the vertical axis (slant) and is analogously
related to the slant angle $\varphi$ to first order by
\begin{equation}
\tan\varphi=\frac{\nabla_{\!h}^{2}\phi_{i}}{2\pi}\frac{(\text{effective grating period})(\text{source--grating distance})}{(\text{detector pixel pitch})^{2}}\,.\label{eq:grating-slant}
\end{equation}

\begin{algorithm}
\begin{algorithmic}[1]
\State ${\phi_{i}}, {y_{ji}}$: input data 
\State $m_{0} \gets \frac{1}{2}$ \Comment upper limit to $\Delta\phi_{ji}, m_0\in(0;1.38]$
\State $\phi_{ji}^{(0)} \gets \phi_{i} \forall j$ \Comment initialization of sampling phases with intended values	
\State $o_{j}^{(0)},a_{j}^{(0)},\phi_{0,j}^{(0)} \gets \underset{o,a,\phi_{0}}{\text{argmin}}\sum_{i}\left(o+a\sin(\phi_{i}^{(0)}-\phi_{0})-y_{ji}\right)^{2}$ \Comment initial sinusoid fits
\For{$k=0\,..\,k_{\max}$}

  \State $\Delta\phi_{ji}^{(k)} \gets \begin{cases}0 & a_j^{(k)}\cos(\phi_{ji}^{(k)}-\phi_{0,j}^{(k)})=0\\\text{softlimit}\left(\frac{(y_{ji}-o_{j}^{(k)})/a_{j}^{(k)}-\sin(\phi_{ji}^{(k)}-\phi_{0,j}^{(k)})}{\cos(\phi_{ji}^{(k)}-\phi_{0,j}^{(k)})},m_0\cos^{2}(\phi_{ji}^{(k)}-\phi_{0,j}^{(k)})\right) & \text{else}\end{cases}$

  \State $w_{ji}^{(k)} \gets a_j^{(k)\,2}\cos^2(\phi_{ji}^{(k)}-\phi_{0,j}^{(k)})$

  \State $\Delta\phi_i^{(k)}, \nabla_{\!h}\phi_i^{(k)}, \nabla_{\!v}\phi_i^{(k)}, \nabla_{\!hv}\phi_i^{(k)}, \nabla^2_{\!h}\phi_i^{(k)} \gets$

  \hfill $ \underset{\Delta\phi_i, \nabla_{\!h}\phi_i, \nabla_{\!v}\phi_i, \nabla_{\!hv}\phi_i, \nabla^2_{\!h}\phi_i}{\mathrm{argmin}} \sum_j w_{ji} \left(\Delta\phi_{i}(j)-\Delta\phi_{ji}^{(k)}\right)^2 $ \Comment for $\Delta\phi_{i}(j)$, cf.\ Eq.~\ref{eq:spatial-phasediff-model}

  \State $\phi_{ji}^{(k+1)} \gets \phi_{ji}^{(k)} + \Delta\phi_i^{(k)}+\nabla_{\!h}\phi_i^{(k)}\,(h-h_{0})+\nabla_{\!v}\phi_i^{(k)}\,(v-v_{0})+\nabla_{\!hv}\phi_i^{(k)}\,(h-h_0)(v-v_{0})+\nabla^2_{\!h}\phi_i^{(k)}\,(h-v_{0})^2 $
  \State $o_{j}^{(k+1)},a_{j}^{(k+1)},\phi_{0,j}^{(k+1)} \gets \underset{o_j,a_j,\phi_{0,j}}{\mathrm{argmin}}\sum_{i}\left(o_j+a_j\sin(\phi_{ji}^{(k+1)}-\phi_{0,j})-y_{ji}\right)^{2}$

\EndFor
.
\end{algorithmic}

\caption{\label{alg:sinusoid-optimization-w-gradient-model}Simultaneous least
squares optimization of abscissa values $\phi_{ji}$ and sinusoid
fits to ordinate samples $y_{ji}$ belonging to independent curves
$j$ sampled at positions $\phi_{ji}=\phi_{i}(j)$ with $\phi_{i}(j)$
being a slowly varying polynomial with respect to the spatial coordinates
$h(j)$ and $v(j)$ accounting for the expected effects due to translations,
magnification and rotations of an interferometer's gratings. The procedure
reduces to Algorithm~\ref{alg:sinusoid-abscissa-optimisation} when
considering only the zeroth order term of $\phi_{i}(j)$.}
\end{algorithm}

\section{Experiment and Results}

\begin{figure}
\centering{}\includegraphics[width=0.7\textwidth]{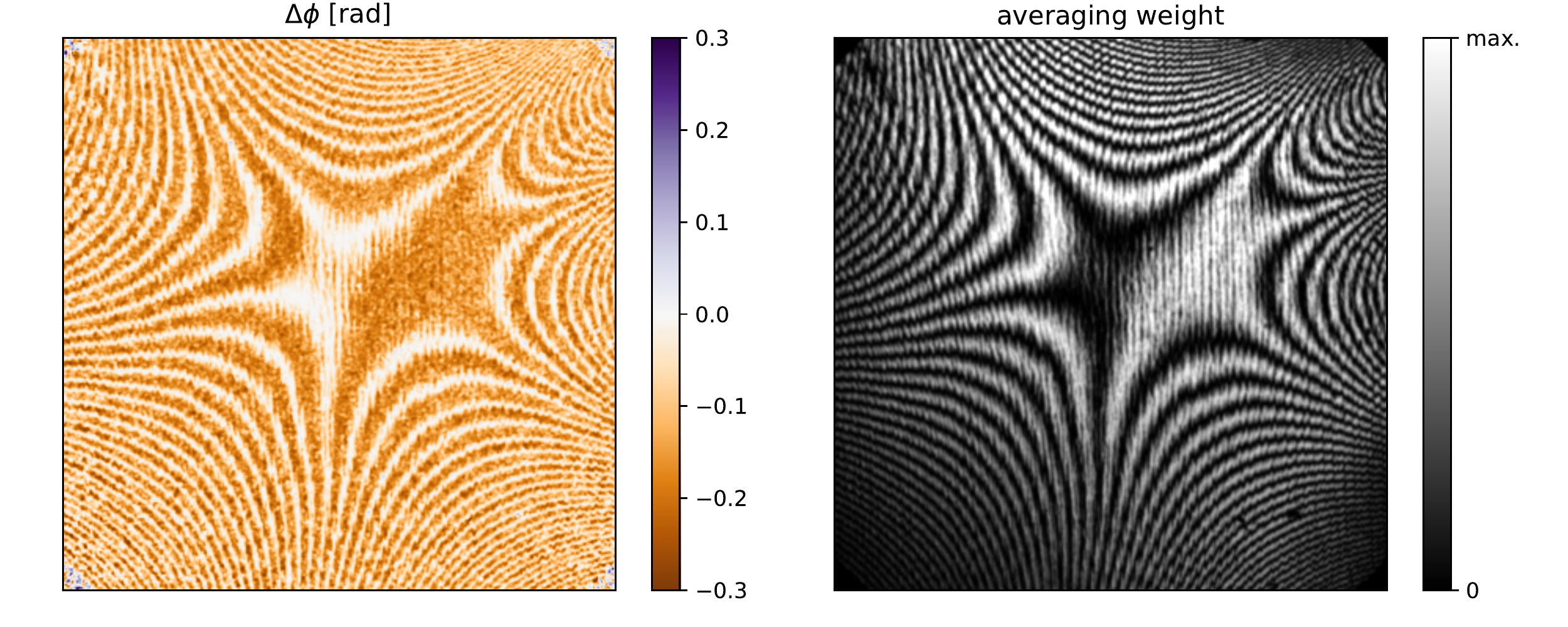}\caption{\label{fig:dphi-and-w-maps}Example of phase differences (left) obtained
from Equation \ref{eq:delta-phi} (cf.\ also Fig.~\ref{fig:delta-phi-demo})
for the first frame of the phase stepping series shown in Fig.~\ref{fig:phase-stepping-series}
with corresponding importance weights (right) as defined by Eq.~\ref{eq:delta-phi-importance-w}.
White regions (0 on the colorbar) in the left hand side $\Delta\phi$
map correspond to samples close to or exactly on turning points of
the fitted sinusoids, where phase differences cannot be effectively
determined. These areas get little weighting in the determination
of the average phase deviation as can be seen by the corresponding
dark fringes in side weighting map on the right. }
\end{figure}
\begin{figure}
\begin{centering}
\par\end{centering}
\centering{}\includegraphics[width=0.9\textwidth]{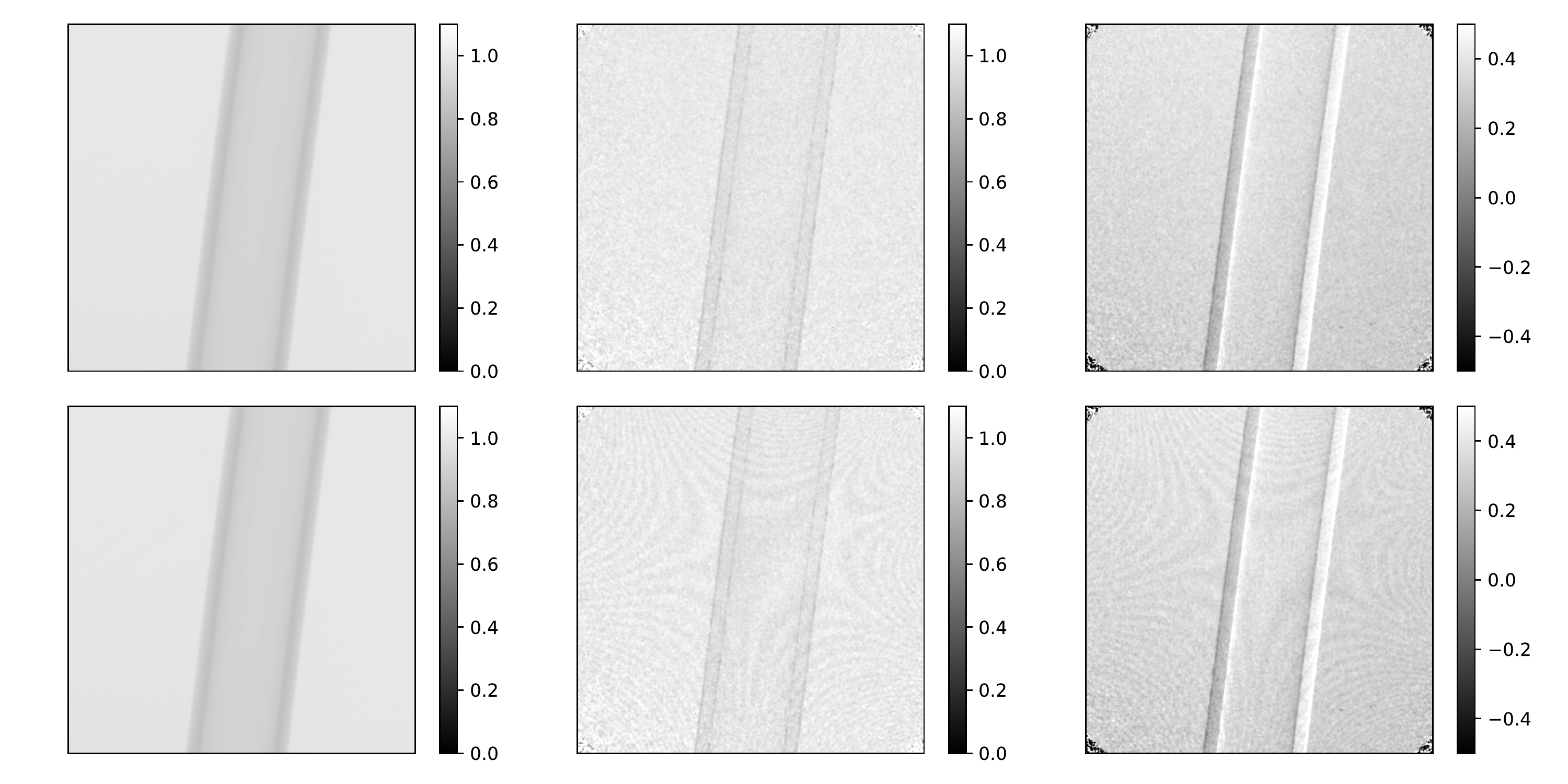}\caption{\label{fig:normalized-images}Transmission, visibility and phase images
(from left to right) of the sample referenced to empty beam images.
The top and bottom rows show results based on phase stepping curve
evaluations with and without correction of the actual sampling positions
respectively. The evaluation based on the assumption of error free
sampling positions (bottom row) exhibits distinctive systematic errors
modulated by the Moiré structure of the reference phase image.}
\end{figure}
\begin{figure}
\centering{}\includegraphics[width=0.45\textwidth]{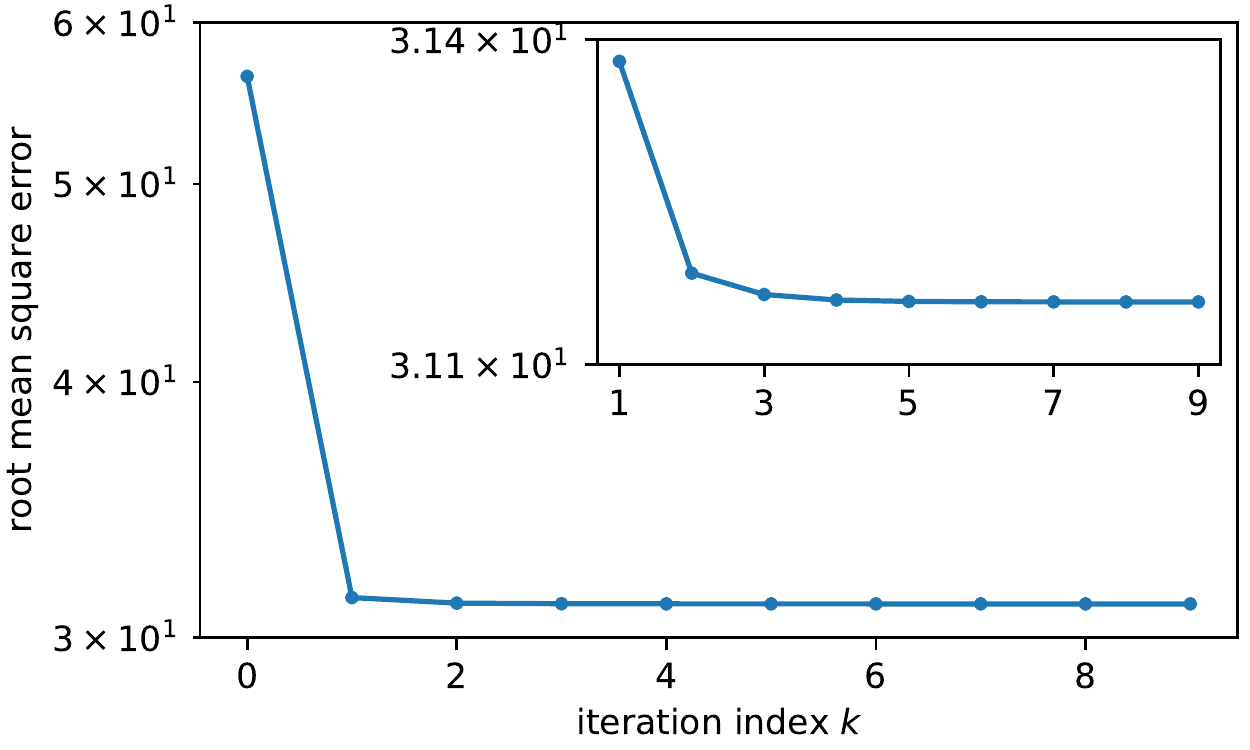}\caption{\label{fig:psc-evaluation-rmse}Root mean square error (RMSE) of the
sinusoid fits to the phase stepping series throughout the iterations
of Algorithm~\ref{alg:sinusoid-abscissa-optimisation}. After the
first correction of the actual sampling locations by Eq.~\ref{eq:delta-phi-average},
the error is reduced by almost a factor of two. The following iterations
further reduce the error confirming the validity of Alg.~\ref{alg:sinusoid-abscissa-optimisation}
for the solution of Eq.~\ref{eq:global-error-min-task}. Additional
consideration of spatially inhomogeneous stepping distances (Eq.~\ref{alg:sinusoid-optimization-w-gradient-model})
further reduces the RMSE by merely 0.1\%. }
\end{figure}

\begin{table}
\centering{}%
\begin{tabular}{ccccc}
\hline 
$\sqrt{\overline{\Delta\phi_{i}^{2}}}$ & $\sqrt{\overline{\left(\nabla_{\!h}\phi_{i}(h-h_{0})\right)^{2}}}$ & $\sqrt{\overline{\left(\nabla_{\!v}\phi_{i}(v-v_{0})\right)^{2}}}$ & $\sqrt{\overline{\left(\nabla_{\!hv}\phi_{i}(h-h_{0})(v-v_{0})\right)^{2}}}$ & $\sqrt{\overline{\left(\nabla_{\!h}^{2}\phi_{i}(h-h_{0})^{2}\right)^{2}}}$\tabularnewline
\hline 
\hline 
$1.2\times10^{-1}$ & $2.4\times10^{-3}$ & $3.6\times10^{-3}$ & $3.6\times10^{-4}$ & $7.5\times10^{-4}$\tabularnewline
\hline 
\end{tabular}\caption{\label{tab:rms-phase-contribs}Root mean square contributions of the
gradient and curvature components of $\Delta\phi_{i}(j)$ to the sampling
phase deviations found for the present phase stepping series in units
of radians. The homogeneous error $\Delta\phi$ is by far the dominating
effect. The contributions of $\nabla_{\!hv}\phi$ and $\nabla_{\!h}^{2}\phi$
range in the order of magnitude of the expected noise level of $10^{-4}\,\text{rad}$
(cf.\ Eqs. \ref{eq:mean-phase-standard-error}, \ref{eq:mean-phase-standard-error-value}).}
\end{table}
\begin{figure}
\begin{centering}
\includegraphics[width=0.9\textwidth]{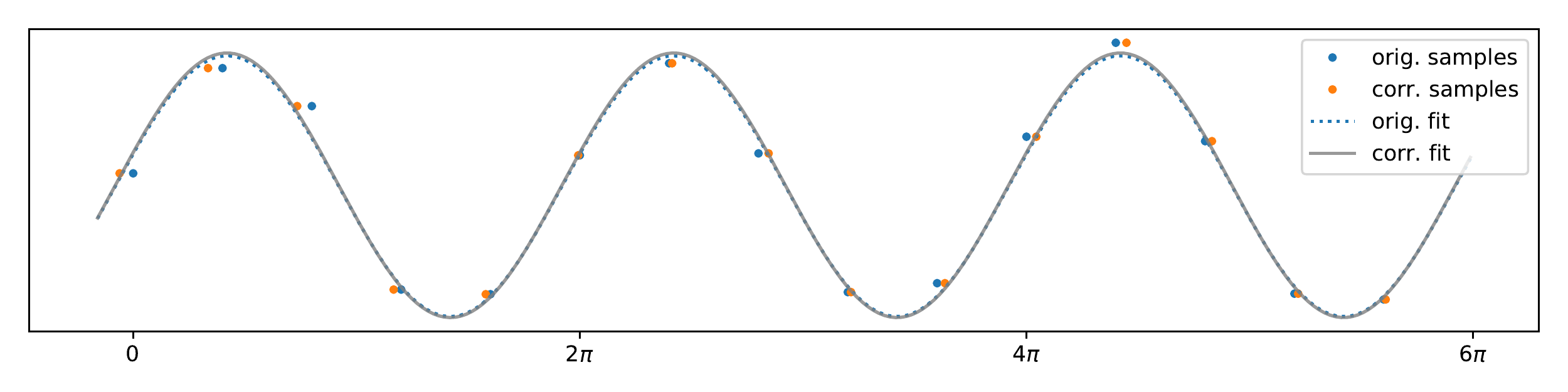}
\par\end{centering}
\centering{}\includegraphics[width=0.9\textwidth]{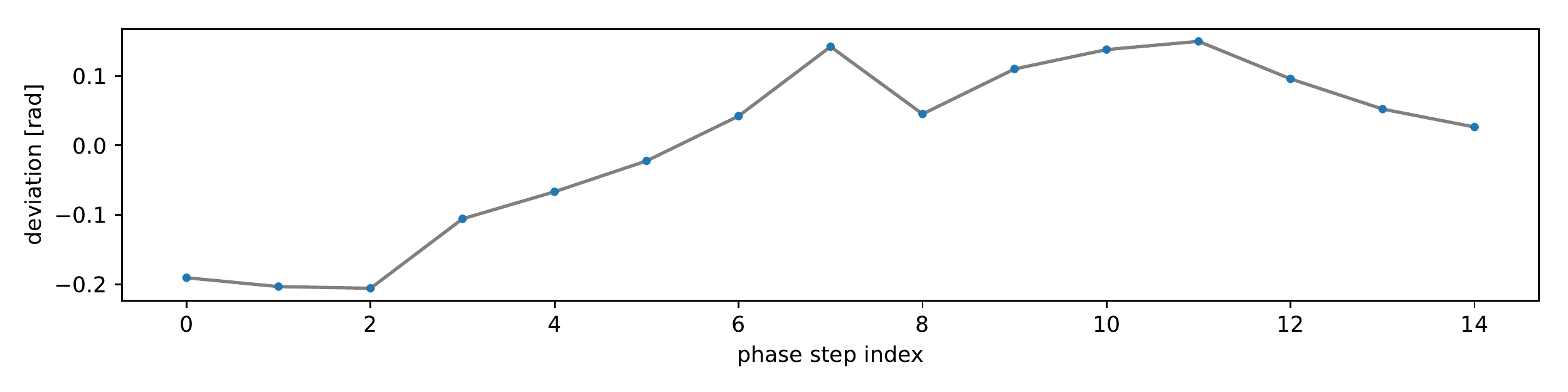}\caption{\label{fig:phase-delta-demo}\emph{Above:} An exemplary phase stepping
curve consisting of 15 steps over three grating periods. Sampled values
are shown both at the originally intended as well as at the inferred
sampling positions (blue and orange markers respectively) along with
the corresponding initial and corrected sinusoid fits. Although the
difference in the resulting fit appears small, it is clearly noticeable
in the final images as shown in Fig.~\ref{fig:normalized-images}.
$\quad$\emph{Below:} Deviations of the phase stepping series' sampling
positions from the intended ones in units of radians. The range of
deviations corresponds to roughly $\pm10\%$ of the intended stepping
increments of $\frac{2}{5}\pi$.}
\end{figure}
Phase stepping series of 15 images sampled at varying relative grating
shifts uniformly distributed over three grating periods have been
acquired both with and without sample in the beam path. Figure~\ref{fig:phase-stepping-series}
shows the first five frames of the empty beam series. The resulting
phase stepping curves at each pixel (indexed by $j$) of the detector
have been evaluated using a least squares fit to a sinusoid model
parameterized by mean $o_{j}$, amplitude $a_{j}$ and phase offset
$\phi_{0,j}$ under the initial assumption of perfectly stepped gratings.
These preliminary results are shown in Figure~\ref{fig:trans-vis-ph-ref}
and correspond to those obtained by classic Fourier analysis of the
phase stepping curves. Deviations of the sampling positions from the
intended ones are then determined based on systematic deviations of
the sampled data from the fitted sinusoids by means of Eq.~\ref{eq:delta-phi-average}
for all 15 frames of the phase stepping series. Figure~\ref{fig:dphi-and-w-maps}
shows an exemplary result for the first frame of the series. Iterating
the sinusoid fits and the corrections to the sampling positions in
order to reduce the overall least squares error (cf.\ Equation~\ref{eq:global-error-min-task})
by means of Algorithm~\ref{alg:sinusoid-abscissa-optimisation},
the sampling positions' deviations are found as shown in Figure~\ref{fig:phase-delta-demo}.
Figure~\ref{fig:normalized-images} shows the reduction of Moiré
modulated systematic errors in the final results, i.e.\ the transmission,
visibility and differential phase images. The root mean square error
is reduced by almost a factor of two in the present example and is
already close to convergence after the first iteration as can be seen
in Fig.~\ref{fig:psc-evaluation-rmse}.

\begin{figure}
\centering{}\includegraphics[width=0.9\textwidth]{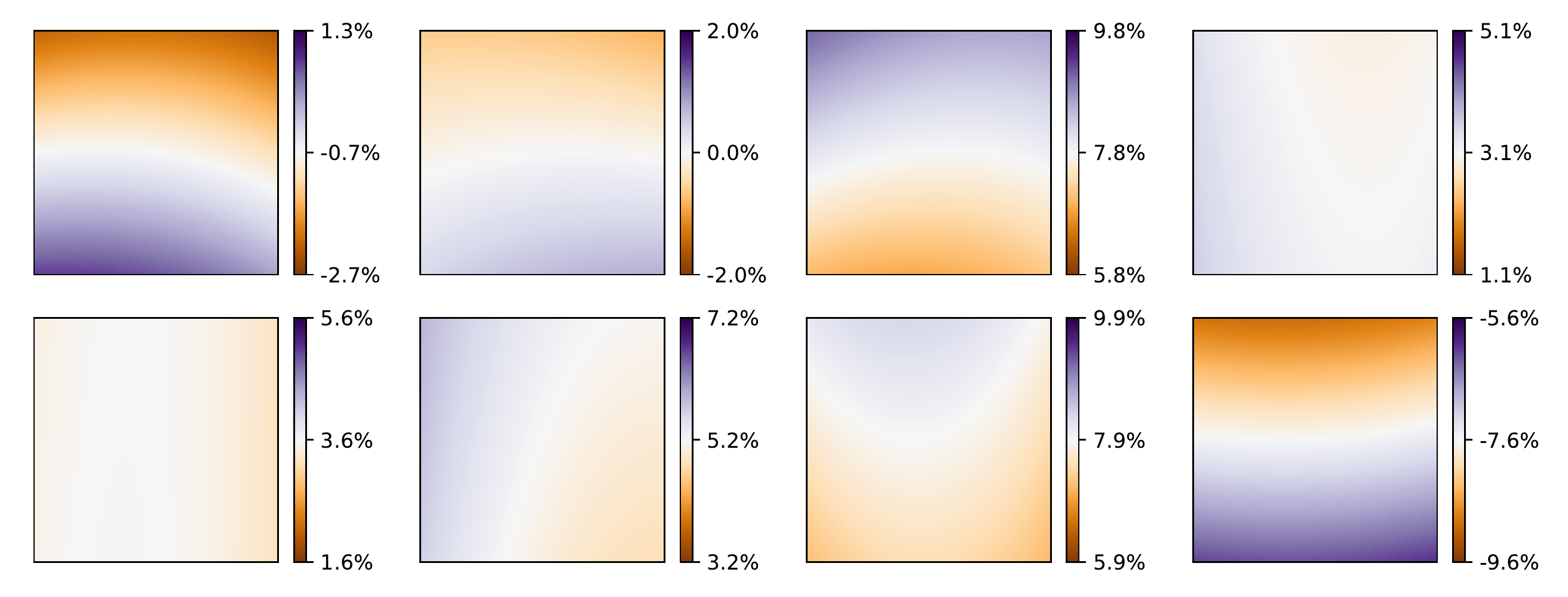}\caption{\label{fig:ddphi-gradients}Relative deviations of the actual phase
steps from the intended step width of $\frac{2\pi}{5}$ between the
first nine frames of the phase stepping series ($\frac{5}{2\pi}(\Delta\phi_{i+1}(j)-\Delta\phi_{i}(j)-\frac{2\pi}{5})$).
The variations $\Delta\phi_{i}(j)$ have been determined by optimization
of Eq.~\ref{eq:global-error-min-w-gradients} assuming the spatial
dependence defined by Eq.~\ref{eq:spatial-phasediff-model} (see
also Algorithm~\ref{alg:sinusoid-optimization-w-gradient-model})}
\end{figure}

In addition to the mean deviations of the phase steps from the intended
positions, spatial gradients throughout the detection area have been
also considered (cf.\ Alg.~\ref{alg:sinusoid-optimization-w-gradient-model}).
Figure~\ref{fig:ddphi-gradients} shows the respective differential
deviations from the intended phase steps between the first 9 frames
of the phase stepping series, normalized to the nominal homogeneous
phase stepping increment of $2\pi/5$. The mean contributions of each
component are listed in Table~\ref{tab:rms-phase-contribs}. While
the homogeneous error of $0.1\,\text{rad}$ ranges within $10\%$
of the nominal step size (or $2\%$ of the grating period), the remaining
effects are two to three orders of magnitude smaller. The root mean
square error of the sinusoid fits for the whole phase stepping series
is reduced by 0.1\% relative to the optimization considering only
homogeneous phase step deviations as shown in Figure~\ref{fig:psc-evaluation-rmse}.
Consequently, the derived images (not shown) are visually equivalent
to those obtained previously (cf.\ Fig.~\ref{fig:normalized-images}). 

\begin{figure}
\centering{}\includegraphics[width=0.9\textwidth]{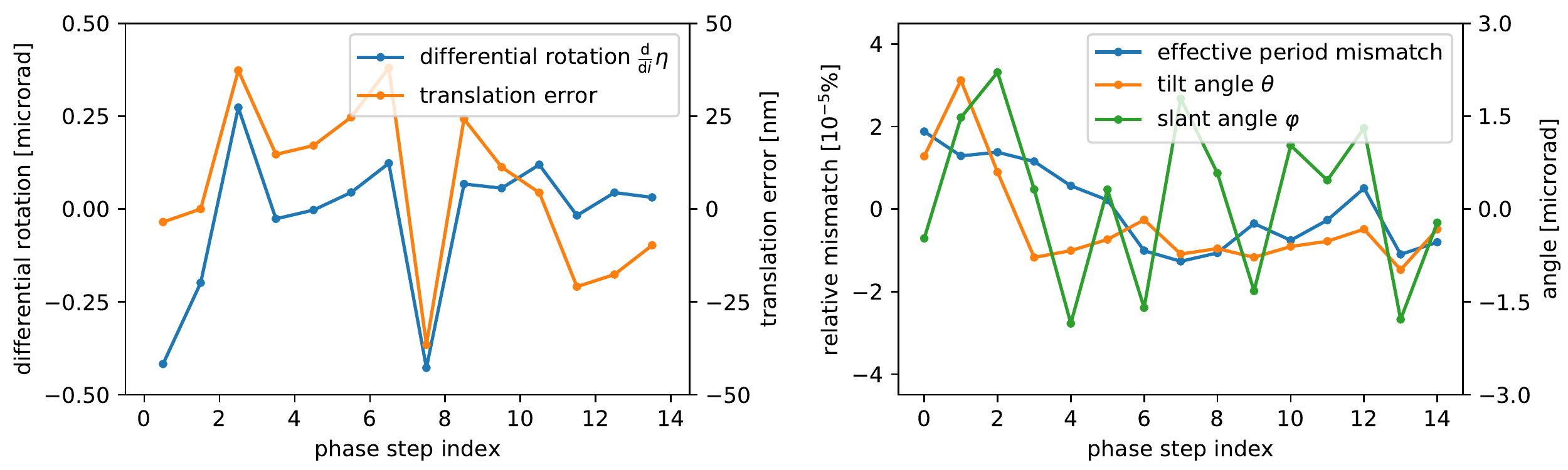}\caption{\label{fig:deviations-interpretations}Quantitative results derivable
from the inhomogeneities in the phase stepping deviations $\Delta\phi_{i}(j)$
(Eq.~\ref{eq:spatial-phasediff-model}). On the left, the change
in tilt angle about the optical axis from frame to frame within the
phase stepping series is shown along with the accompanying linear
motion error. Rotation correlated translations indicate an off-center
pivot point. On the right hand side, the relative grating scaling
error is shown along with the found tilt and slant about the horizontal
and vertical axis. These quantities represent deviations from the
mean grating alignment throughout the phase stepping series. The tilt
and slant angles range close to the expected noise level (cf.\ Table~\ref{tab:rms-phase-contribs}
and Eq.~\ref{eq:mean-phase-standard-error-value}).}
\end{figure}

Figure~\ref{fig:deviations-interpretations} shows variations in
the relative alignment of the gratings derived from the inhomogeneous
phase stepping analysis by means of Equations \ref{eq:grating-rotation}\textendash \ref{eq:grating-slant}.
Besides deviations from the nominal linear motion of the gratings,
minute rotations as well as subtle changes in relative magnification
can also be detected. The correlation between grating rotations and
translational errors visible in the left hand side graph in Figure~\ref{fig:deviations-interpretations}
indicate rotations about an off-center pivot point about $10^{-1}\text{m}$
below the grating center, which is consistent with the actual placement
of the phase stepping actuator in the experimental setup. Although
the analog correlation between tilts about the horizontal axis and
translation (along the optical axis) induced variations in magnification
is much less pronounced (Fig.~\ref{fig:deviations-interpretations},
right hand side), the mean trend and magnitude are also consistent
with the assumption of a pivot point below the field of view. However,
the observed magnitude ($10^{-7}$) of the relative mismatch of the
effective grating periods is as well explicable by temperature variations
in the order of magnitude of $10^{-1}\text{K}$ given a thermal expansion
coefficient in the order of magnitude of $10^{-6}\text{K}^{-1}$ for
the typical wafer materials silicon and graphite. Finally, the phase
stepping inhomogeneities further suggest rotational motions about
the vertical axis on the microrad scale (also shown in Fig.~\ref{fig:deviations-interpretations}).

As a crude error assessment, the standard error of the mean phase
deviation can be estimated from the sinusoid fits' root mean square
error:
\begin{equation}
\begin{alignedat}{1}\sigma_{\mathrm{mean}} & \approx\frac{1}{\sqrt{\text{contributing detector pixels}}}\,\frac{\text{sinusoid fit RMSE}}{\text{mean sinusoid amplitude}}\\
 & \approx\sqrt{\frac{2}{\text{detector pixels}}}\,\frac{\text{sinusoid fit RMSE}}{\text{mean sinusoid amplitude}}\,.
\end{alignedat}
\label{eq:mean-phase-standard-error}
\end{equation}
The latter corresponds for the present data set to about $6\%$ of
the mean observed sinusoid amplitude, which directly translates to
$6\times10^{-2}\,\text{rad}$ on the abscissa. Given the amount of
detector pixels contributing to the least squares fits of $\Delta\phi_{i}(j)$
within each frame of the phase stepping series, a standard error in
the order of magnitude of
\begin{equation}
\text{\ensuremath{\sigma_{\mathrm{mean}}}}\approx10^{-4}\,\text{rad}\label{eq:mean-phase-standard-error-value}
\end{equation}
 results. This implies that, according to the results given in Table~\ref{tab:rms-phase-contribs},
the tilt and slant contributions $\nabla_{\!hv}\phi$ and $\nabla_{\!h}^{2}\phi$
are close to the expected noise level for the present case.


\section{Discussion and Conclusion}

A fast converging iterative algorithm for the joint optimization
of both the sinusoid model parameters and the actual sampling locations
for the evaluation of grating interferometric phase stepping series
has been proposed. Within the optimization procedure, spatially varying
phase stepping increments due to further mechanical degrees of freedom
besides the intended translatory stepping motion can also be accounted
for. Mean phase stepping errors of up to 10\% of the nominal step
width have been found for a typical data set, and their correction
results both in a considerable reduction of the overall root mean
square error by almost a factor of two as well as a significant visual
improvement of the final images. Although higher order effects are
observable, their contribution was found to be two to three orders
of magnitude smaller than that of the mean stepping error, and their
correction thus did not contribute to further improvements in visual
image quality in the present case. However, the higher order deviations
allow the detection of minute motions of the gratings and thus provide
a valuable tool for the monitoring and debugging of experimental setups.
First order approximations for the relations between spatial phase
variations and mechanical degrees of freedom of the moved grating
have been given. For the present data set, linear motion errors up
to $25\,\text{nm}$ as well as rotational motions on the microrad
scale have been inferred from the phase stepping series. While the
tilt and slant angles about the horizontal and vertical axes respectively
have been found to range close to the expected noise level and should
rather be interpreted as upper limits to actual motions, magnification
changes in the range of $10^{-7}$ and sub-microrad rotations about
the optical axis were well detectable. The expected correlations between
rotation and translation due to an off-center pivot point further
support the plausibility of the results. Especially the crosstalk
between sub-microrad rotations and effective translations indicates
that noticeable phase stepping errors will be almost inevitable even
for very carefully designed experiments, wherefore an optimization
based evaluation of the phase stepping series as proposed in Algorithm~\ref{alg:sinusoid-abscissa-optimisation}
is generally advisable. With processing speeds in the range of $0.1\,\text{s}$
per phase stepping series, it is well suitable as a standard processing
method also for large image series.


\begin{thebibliography}{1}
\bibitem{David2002}C. David, B. Nöhammer, H. H. Solak, E. Ziegler:
Differential x-ray phase contrast imaging using a shearing interferometer.
Appl. Phys. Lett. 81 (17) 3287\textendash 3289 (2002). doi: 10.1063/1.1516611 

\bibitem{Momose2003}Atsushi Momose, Shinya Kawamoto , Ichiro Koyama,
Yoshitaka Hamaishi , Kengo Takai and Yoshio Suzuki: Demonstration
of X-Ray Talbot Interferometry. Jpn. J. Appl. Phys. Vol. 42 (2003)
pp. L 866\textendash L 868 

\bibitem{Pfeiffer2006}F. Pfeiffer, T. Weitkamp, O. Bunk, C. David:
Phase retrieval and differential phase-contrast imaging with low-brilliance
X-ray sources. Nature physics, 2(4), 258\textendash 261 (2006), doi:10.1038/nphys2

\bibitem{Revol2010}V. Revol, C. Kottler, R. Kaufmann, U. Straumann,
C. Urban: Noise analysis of grating-based x-ray differential phase
contrast imaging. Rev. Sci. Instrum. 81, 073709 (2010), doi:10.1063/1.3465334

\bibitem{Seifert2016}M. Seifert, S. Kaeppler, C. Hauke, F. Horn,
G. Pelzer, J. Rieger, T. Michel, C. Riess, G. Anton: Optimisation
of image reconstruction for phase-contrast x-ray Talbot\textendash Lau
imaging with regard to mechanical robustness. Phys. Med. Biol. 61,
6441 (2016), doi:10.1088/0031-9155/61/17/6441

\bibitem{Vargas2013}J. Vargas, C.O.S. Sorzano, J.C. Estrada, J.M.
Carazo: Generalization of the Principal Component Analysis algorithm
for interferometry. Opt. Comm. 286, 130\textendash 134 (2013)
\end{thebibliography}
\end{document}